\documentclass[runningheads]{llncs}
\usepackage{graphicx}
\usepackage{subcaption, amsmath, soul, color, calrsfs, array,hyperref}
\usepackage[misc]{ifsym}
\DeclareMathAlphabet{\pazocal}{OMS}{zplm}{m}{n}
\newcommand{\Da}{\mathcal{D}}

\begin{document}
%
\title{Deep Sequential Mosaicking of Fetoscopic Videos}
%
%
\author{Sophia Bano\inst{1}($^{\textrm{\Letter}}$) \and
Francisco Vasconcelos\inst{1} \and
Marcel Tella Amo\inst{1} \and
George Dwyer\inst{1} \and
Caspar Gruijthuijsen\inst{2} \and
Jan Deprest\inst{4} \and
Sebastien Ourselin\inst{3} \and
Emmanuel Vander Poorten\inst{2} \and
Tom Vercauteren\inst{3} \and
Danail Stoyanov\inst{1}}
\authorrunning{S. Bano et al.}

\institute{Wellcome/EPSRC Centre for Interventional and Surgical Sciences(WEISS) and Department of Computer Science, University College London, London, UK
\email{sophia.bano@ucl.ac.uk} \and
Department of Mechanical Engineering, KU Leuven University, Leuven, Belgium  \and
School of Biomedical Engineering and Imaging Sciences, King\textsc{\char13}s College London, London, UK  \and
Department of Development and Regeneration, University Hospital Leuven, Leuven, Belgium}

\maketitle              
\begin{abstract}
Twin-to-twin transfusion syndrome treatment requires fetoscopic laser photocoagulation of placental vascular anastomoses to regulate blood flow to both fetuses. Limited field-of-view (FoV) and low visual quality during fetoscopy make it challenging to identify all vascular connections. Mosaicking can align multiple overlapping images to generate an image with increased FoV, however, existing techniques apply poorly to fetoscopy due to the low visual quality, texture paucity, and hence fail in longer sequences due to the drift accumulated over time. Deep learning techniques can facilitate in overcoming these challenges. Therefore, we present a new generalized Deep Sequential Mosaicking (DSM) framework for fetoscopic videos captured from different settings such as simulation, phantom, and real environments. DSM extends an existing deep image-based homography model to sequential data by proposing controlled data augmentation and outlier rejection methods. Unlike existing methods, DSM can handle visual variations due to specular highlights and reflection across adjacent frames, hence reducing the accumulated drift. We perform experimental validation and comparison using 5 diverse fetoscopic videos to demonstrate the robustness of our framework.

\end{abstract}
\section{Introduction}
Twin-to-twin transfusion syndrome (TTTS) can occur during identical twin pregnancies where abnormal vascular anastomoses in the monochorionic placenta result in uneven blood flow between the fetuses~\cite{baschat2011twin}. Fetoscopic laser photocoagulation is the most effective treatment for regulating the blood flow. During treatment, the clinician first visually explores the placenta using fetoscopic video to identify vascular anastomoses, building a mental map and treatment plan. Limited FoV, poor visibility and limited maneuverability of the fetoscope introduce challenges that increase procedural time, can lead to complications and impede verifying completion~\cite{vasconcelos2018towards}. Mosaicking can align multiple overlapping images to generate an image with increased FoV. Hence it can provide computer-assisted intervention support to ease the localization of the vascular anastomoses sites. 

Mosaicking has recently gained attention to increase the FoV in fetoscopy \cite{reeff2006mosaicing,daga2016real,tella2018probabilistic,tella2019pruning,peter2018retrieval}. Totz et al.~\cite{totz2011dense} presented a dynamic view expansion and surface reconstruction approach for minimally invasive surgery by analyzing stereo laparoscopy videos. Reeff at al.~\cite{reeff2006mosaicing} and Daga et al.~\cite{daga2016real} utilized a classical image feature-based matching method for creating mosaics from planar placenta images. The relative transformations between pairs of consecutive fetoscopic images are computed and combined in a chain with respect to a reference frame to generate the mosaic. Error in the relative transformations can propagate to introduce large drift in the overall mosaic, where an electromagnetic tracker (EMT) can be integrated within the fetoscope to minimize any drifting errors~\cite{tella2018probabilistic}. However, integrating an EMT sensor with a fetoscope in-vivo is still an open challenge due to limited form-factor of the fetoscope and due to regulation. To this end, an existing registration technique avoided explicit feature correspondence by utilizing pixel-wise alignment of gradient orientations for a single in-vivo fetoscopic video~\cite{peter2018retrieval}. Fetoscopic videos are captured from monocular cameras and pose challenges for mosaicking due to varying visual quality due to various types of fetoscopes, occlusions, specular highlights, lack of visual texture, poor visibility due to turbid amniotic fluid and non-planar views~\cite{gaisser2018stable}.

Recently, deep image homography estimation methods have been proposed \cite{detone2016deep,nguyen2018unsupervised} that estimate the homography between pairs of image patches extracted from an image. We observe that a full mosaic is generated by computing sequential homographies between adjacent frames, where a fetoscopic video poses challenges such as specular reflections, amniotic fluid particles, and occlusions. This affects the stitching problem, however, such challenges can be tackled when the homography is estimated using multiple pairs of image patches extracted at random from adjacent frames. In this paper, we employ this approach and propose the first generalized Deep Sequential Mosaicking (DSM) framework for creating mosaics with minimum drift from long-range fetoscopic videos captured from various fetoscopes. We adopt the deep image-based homography estimation method~\cite{detone2016deep} to incorporate sequential data by proposing the Controlled Data Augmentation (CDA) and outlier rejection methods. CDA assumes that the transformation between two adjacent frames contains rotation and translation only, and uses a small set of fetoscopic images of varying quality and appearance, for training. To eliminate the error due to varying visual quality and texture paucity between adjacent frames, we propose the outlier rejection method. This increases the robustness by pruning patch-based homography estimates between adjacent frames. CDA along with the outlier rejection minimize the drift without the use of any external sensors and generate reliable mosaics in this challenging application. Comparison with existing methods and validation on 5 datasets verifies the promising generalization capabilities of our method.

\section{Homography Estimation with Deep Learning}
\label{sec:deep_homo}
\begin{figure}[!t]
	\centering
	\includegraphics[width=0.8\textwidth]{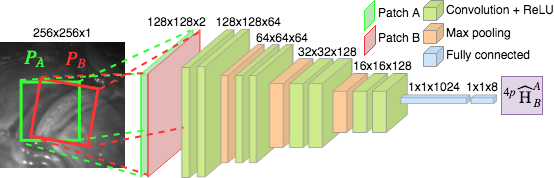}	
	\caption{Deep image homography network with controlled data augmentation.}
	\label{fig:HomoNet_arch}       
\end{figure}

The Deep Image Homography (DIH) model~\cite{detone2016deep} estimates the relative homography between pairs of image patches extracted from a single image. This model uses the 4-point homography parameterization $^{4p}\mathbf{H}$, instead of the 3$\times$3 parameterization $\mathbf{H}$, as the rotation and shear components in $\mathbf{H}$ have smaller magnitude compared to the translation, thus have a small effect on the training loss. Let $(u_i,v_i)$ and $(u_i',v_i')$ denote the four corners of image patch $P_A$ and $P_B$. Then the 4-point homography $^{4p}\mathbf{H}$ is given by:
\begin{equation}
^{4p}\mathbf{H} = \begin{bmatrix}
\Delta u_1 &\Delta u_2 &\Delta u_3 &\Delta u_4\\ 
\Delta v_1 &\Delta v_2 &\Delta v_3 &\Delta v_4
\end{bmatrix}^{T},\,\,\,
\begin{matrix}
\mbox{where}\,\,\,\Delta u_i = u_i' - u_i,\,\,\Delta v_i = v_i' - v_i \\
\mbox{and}\,\, i =1,2,3,4\,\,\,\,\,\,\,\,\,\,\,\,\,\,\,
\end{matrix}
\end{equation}
DIH~\cite{detone2016deep} uses a VGG-like architecture, with 8 convolutional and 2 fully connected layers (Fig.~\ref{fig:HomoNet_arch}). The input of the network is P$_A$ and P$_B$, and output is their relative homography. 
Note that~\cite{detone2016deep} used the MS-COCO dataset for training, where pair of patches were extracted from a single real image, free of artifacts~(e.g.~specular highlights, amniotic fluid particles) that appear in sequential data.   

DIH~\cite{detone2016deep} generated the training data by randomly selecting P$_A$ from a grayscale image and randomly perturbing its corners to obtain P$_B$ and the Ground-Truth (GT). We observe through experimentation that such data augmentation results in scenarios that are challenging for the network to learn, hence results in a large error (Fig.~\ref{fig:syn_visualization}(d) and Fig.~\ref{fig:rmse_photo}). While such errors are acceptable in image-based homography~\cite{detone2016deep}, for mosaicking even a small error in pairwise homography accumulates over time resulting in increased drift. Therefore, this data generation approach cannot be used as it is for sequential mosaicking. 
\section{Deep Sequential Mosaicking (DSM)}
\begin{figure}[!t]
	\centering
	\includegraphics[width=1.0\textwidth]{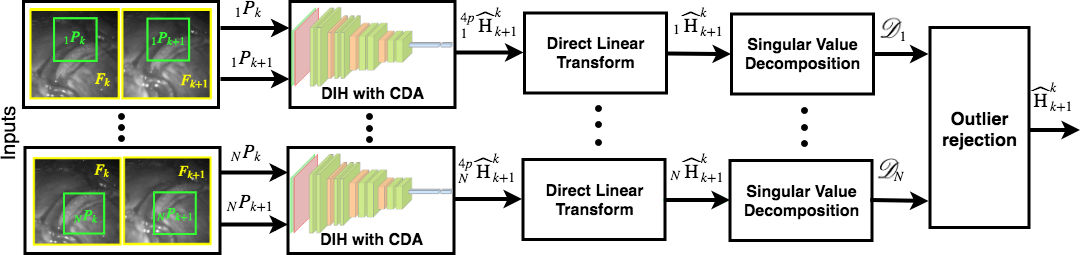}
	\caption{Overview of the proposed Deep Sequential Mosaicking( DSM) method that uses Controlled Data Augmentation (CDA) for training Deep Image Homography (DIH) model and outlier rejection for pruning the homography estimates.\looseness=-1}
	\label{fig:block_diagram}      
\end{figure}

Mosaic from an image sequence can be generated by finding the pairwise homographies between adjacent frames, followed by computing the relative homographies with respect to a reference frame. The GT pairwise homographies are unknown in fetoscopic videos since they are captured from a monocular camera. Therefore, only through visualization, we can observe the error accumulated over time. For minimizing this error, in our proposed DSM, the relative homography is learned between patches that are extracted from a single image following the CDA (Sec.~\ref{sec:data_gen}). Unlike \cite{detone2016deep}, in practice homography is computed between two adjacent frames, having specular highlights and lack of texture, in fetoscopic videos. Therefore, testing by using pairs of patches from two adjacent frames results in varying $\mathbf{H}$. To overcome this error, we propose an outlier rejection step (Sec.~\ref{sec:post-processing}) to improve the estimation. 
During testing (Fig.~\ref{fig:block_diagram}), we compute homographies between pairs of adjacent frames $N$ times by randomly selecting the location of P$_A$. The estimated $^{4p}\mathbf{H}$ is converted to $\mathbf{H}$ by applying Direct Linear Transform (DLT), followed by its decomposition using Singular Value Decomposition (SVD) and outlier rejection for removing inaccurate estimations.  

%
%
\subsection{Controlled Data Augmentation (CDA)}
\label{sec:data_gen}
Pairwise homography between two consecutive frames F$_k$ and F$_{k+1}$ are related by affine transformations including rotation, translation, scale, and shear. A TTTS procedure is performed at a fixed distance from the placenta, hence the scale remains constant. Fetoscope motion is physically constrained by the incision point (remote center of motion), which makes shear very small in consecutive frames, compared to rotation and translation. Therefore, we neglect the scale and shear components and assume that F$_k$ and F$_{k+1}$ are related by translation and rotation only. This helps to minimize the error in relative homography and consequently reduce the drift in mosaicking. 
For CDA, given a grayscale image I, an image patch P$_A$ is extracted at a random location with corner points $(u_i, v_i)$, where $i = 1, 2, 3, 4$. 
Rotation by $\beta$ and translation by $(d_x,d_y)$ is applied:
\begin{equation}
\begin{bmatrix}
u_i'\\ 
v_i'
\end{bmatrix}
=
\begin{bmatrix}
cos\beta & sin \beta\\ 
-sin \beta & cos \beta 
\end{bmatrix}
\begin{bmatrix}
u_i\\ 
v_i
\end{bmatrix}
+
\begin{bmatrix}
d_x\\ 
d_y
\end{bmatrix},
\end{equation}
to obtain P$_B$, where $\beta$, $d_x$ and $d_y$ are empirically selected. During training, the relative homography is learned between patches that are extracted from a single image following the CDA. Due to lack of texture and poor contrast in fetoscopic videos, homography between two consecutive frames may not be accurate.
\subsection{Homography Matrix Decomposition and Outlier Rejection}
\label{sec:post-processing}
To obtain the most consistent homography matrix, we first decompose the homography matrix by applying SVD~\cite{malis2007deeper}:
\begin{equation}
\begin{bmatrix}
\widehat{h}_{11} & \widehat{h}_{12} \\ 
\widehat{h}_{21} & \widehat{h}_{22} 
\end{bmatrix}
=
\begin{bmatrix}
cos\widehat{\theta} & sin \widehat{\theta} \\ 
-sin \widehat{\theta} & cos \widehat{\theta} 
\end{bmatrix}
\begin{bmatrix}
\widehat{s}_g & 0 \\ 
0 & \widehat{s}_h 
\end{bmatrix}
\begin{bmatrix}
cos\widehat{\gamma} & sin \widehat{\gamma} \\ 
-sin \widehat{\gamma} & cos \widehat{\gamma} 
\end{bmatrix},
\label{eq:svd}
\end{equation} 
and $\widehat{t}_x = \widehat{h}_{13}$, $\widehat{t}_y= \widehat{h}_{23}$ are the translation components. By solving eq.~\ref{eq:svd}, we obtain the decomposed parameters, $\Da =\{\widehat{\theta},\, \widehat{\gamma},\, \widehat{s}_g,\, \widehat{s}_h\}$~\cite{malis2007deeper}. Next, for F$_k$ and F$_{k+1}$, we compute $_n\widehat{\mathbf{H}}_{k+1}^{k}$ for $N=99$ iterations by selecting a new random patch pair $_n$P$_k$ and $_n$P$_{k+1}$ at each iteration and obtain $N$ decompose parameters, represented for example as $(\widehat{\theta}_n)_{n=1}^{N}$. 
The variations in $(\widehat{s}_{gn})_{n=1}^{N}$ and $(\widehat{s}_{hn})_{n=1}^{N}$  are very small due to fixed scale assumption, but are significant in $(\widehat{\theta}_n)_{n=1}^{N}$ and $(\widehat{\gamma}_n)_{n=1}^{N}$. 
Since the first and third matrices in eq.~\ref{eq:svd} are orthogonal, $\widehat{\theta}_n = - \widehat{\gamma}_n$, filtering either of the two has the same effect. 
We apply median filtering, since it is useful for mitigating the effect of the outliers, to $(\widehat{\theta}_n)_{n=1}^{N}$ to get its argument $i$, giving the most consistent value for $\theta$. This argument is used to obtain $\widehat{\gamma}_i$, $\widehat{s}_{xi}$, $\widehat{s}_{yi}$, $\widehat{t}_{xi}$ and $\widehat{t}_{yi}$, that are then plugged into eq.~\ref{eq:svd} to get the consistent $_i\widehat{\mathbf{H}}_{k+1}^{k}$.

\begin{table}[t!]
	\centering
	\scriptsize
	\resizebox{1\columnwidth}{!}{%
	\begin{tabular}{|m{2.0cm}||m{2.4cm}|m{2.4cm}|m{2.4cm}|m{2.4cm}|m{2.4cm}|}
	\hline
	Representative frame &\parbox[c]{2.4cm}{\centering \includegraphics[width=1.7cm]{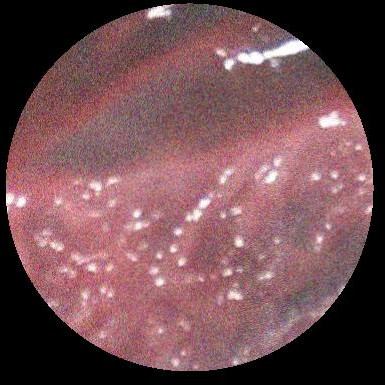}} 
	    &\parbox[c]{2.4cm}{\centering	\includegraphics[width=1.7cm]{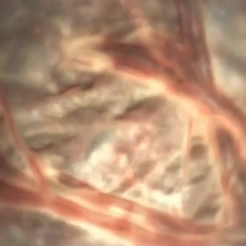}} &\parbox[c]{2.4cm}{\centering \includegraphics[width=2.1cm]{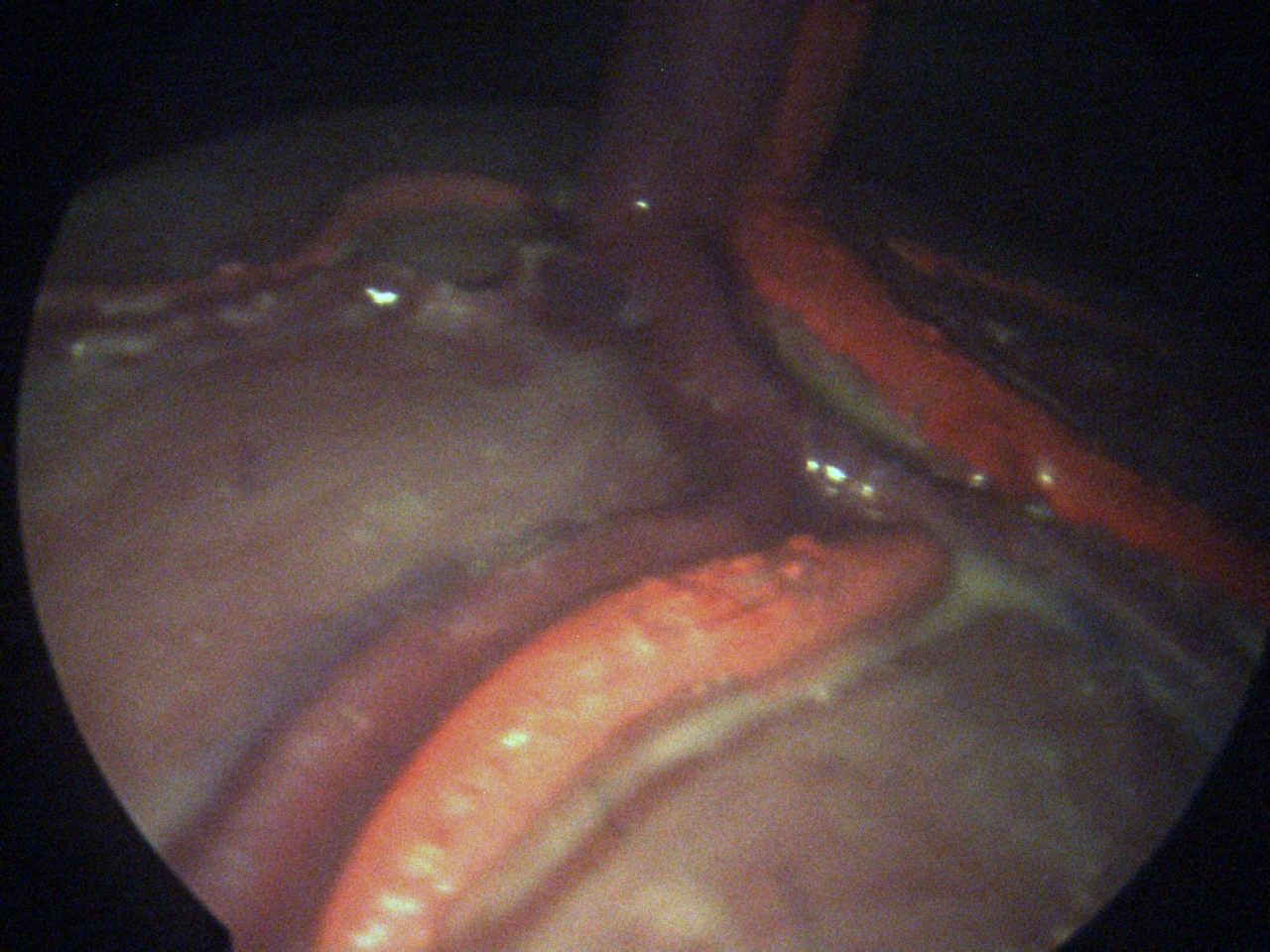}} 
	    &\parbox[c]{2.4cm}{\centering \includegraphics[width=1.7cm]{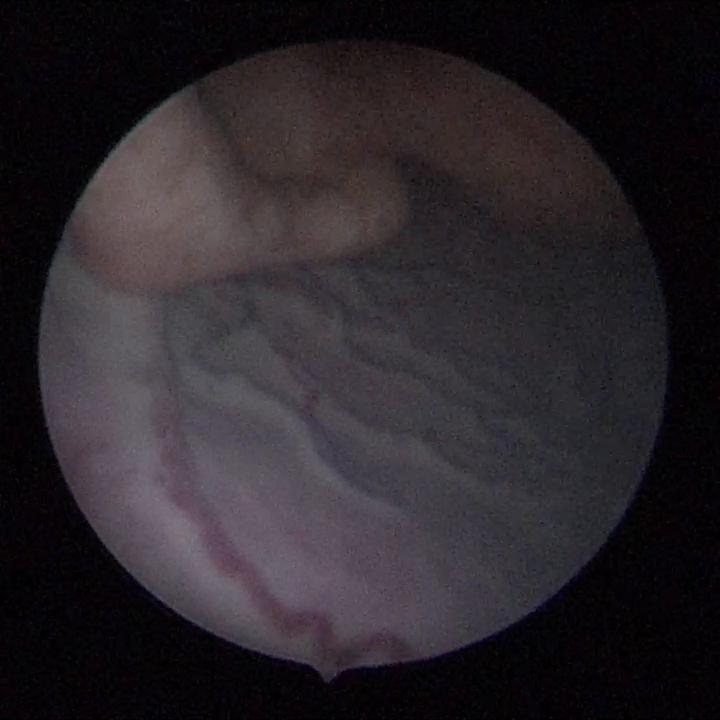}} 
	    &\parbox[c]{2.4cm}{\centering \includegraphics[width=1.7cm]{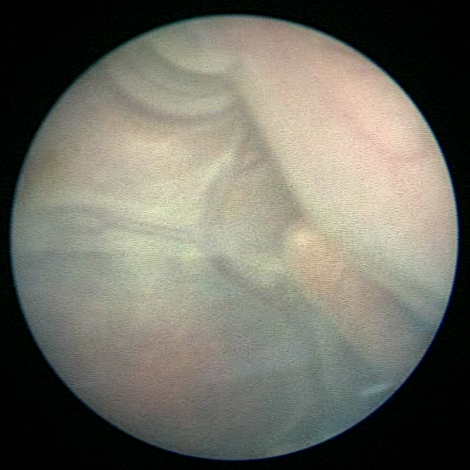}}	\\
	    \hline
	    & & & & & \\[-0.9em]  
	Data type &Synthetic (SYN) &\parbox[c]{2.4cm}{Ex-vivo in water \\ (EX)} &\parbox[c]{2.4cm}{Phantom without \\fetus (PHN1)} &\parbox[c]{2.4cm}{TTTS Phantom in \\water (PHN2)} &\parbox[c]{2.4cm}{Invivo TTTS \\procedure (INVI)} \\	
	 & & & & & \\[-0.9em]  
	\hline 		
	 & & & & & \\[-0.9em]  
	Imaging source &\multicolumn{1}{c|}{-} &\multicolumn{1}{c|}{Stereo} & \multicolumn{1}{c|}{Rigid $30^{\circ}$ scope} &\multicolumn{1}{c|}{Rigid scope} &\multicolumn{1}{c|}{Rigid scope} \\
	 & & & & & \\[-0.9em]  
	\hline 		
	 & & & & & \\[-0.9em]  
	No. of frames &\multicolumn{1}{c|}{811} &\multicolumn{1}{c|}{404} & \multicolumn{1}{c|}{681} &\multicolumn{1}{c|}{400} &\multicolumn{1}{c|}{200} \\
	 & & & & & \\[-0.9em]  
	\hline
	 & & & & & \\[-0.9em]  
	\parbox[c]{2.0cm}{Resolution \\ (pixels)} &\multicolumn{1}{c|}{385 $\times$ 385} &\multicolumn{1}{c|}{250 $\times$ 250} & \multicolumn{1}{c|}{1280 $\times$ 960} &\multicolumn{1}{c|}{720 $\times$ 720} &\multicolumn{1}{c|}{470 $\times$ 470} \\
	 & & & & & \\[-0.9em]  
	\hline
	 & & & & & \\[-0.9em]  
	\parbox[c]{2.0cm}{Crop resolution \\ (pixels)} &\multicolumn{1}{c|}{260 $\times$ 260} &\multicolumn{1}{c|}{250 $\times$ 250} & \multicolumn{1}{c|}{834 $\times$ 834} &\multicolumn{1}{c|}{442 $\times$ 442} &\multicolumn{1}{c|}{312 $\times$ 312} \\
	 & & & & & \\[-0.9em]  
	\hline
	 & & & & & \\[-0.9em]  
	Camera view &\multicolumn{1}{c|}{Planar} &\multicolumn{1}{c|}{Planar} & \multicolumn{1}{c|}{Non-planar} &\parbox[c]{2.4cm}{Non-planar \\heavy occlusions} &\parbox[c]{2.4cm}{Non-planar \\heavy occlusions} \\
	 & & & & & \\[-0.9em]  
	\hline	
	 Motion type &\multicolumn{1}{c|}{Circular} &\multicolumn{1}{c|}{Spiral} &\multicolumn{1}{c|}{Circular freehand} &\parbox[c]{2.4cm}{Exploratory \\freehand} &\parbox[c]{2.4cm}{Exploratory \\freehand} \\ [4pt]
	\hline	
	\end{tabular}
	}
	\caption{Main characteristics of the datasets used for the experimental analysis.}
	\label{tab:dataset}
\end{table}
\section{Experimental Setup and Evaluation Protocol}
For experimental analysis, we use 5 fetoscopic videos (Table.~\ref{tab:dataset}), which include a synthetic video (SYN) - a discontinuous version of this sequence was used in \cite{tella2018probabilistic}, an ex-vivo in water (EX) data reported in \cite{dwyer2017continuum}, a placenta phantom (PHN1), a TTTS phantom\footnote{\scriptsize TTTS phantom from Surgical Touch Simulator: \url{https://www.surgicaltouch.com/}} in water (PHN2) depicting an in-vivo procedure and an in-vivo TTTS procedure (INVI). Note from Table.~\ref{tab:dataset} the variability in visual quality, appearance, resolution, imaging source, camera views and captured motion. These variations pose challenging scenarios for mosaicking methods.

For training, we use 600 frames extracted at random from SYN, PHN1, PHN2, INVI and another ex-vivo still images dataset (not used in testing as it is not a video sequence). EX~(Table~\ref{tab:dataset}) is not use during training, hence it is an unseen data for testing. We extract square frames, from the circular FoV of fetoscopic videos, to be used as the input to DSM. All images are converted to grayscale and resized to $256 \times 256$ pixels. 
We use Keras with Tensorflow backend for the implementation and train our network for about 15 hours on a Tesla V100 (32GB) using learning rate of $10^{-4}$ and ADAM optimizer. DIH with CDA is trained for 60,000 epochs with a batch size of $32$. In each epoch, pairs of patches are generated by randomly selecting $\beta$ between~$(-5,+5)$ \textit{degrees}, and $d_x$ and $d_y$ between $(-16,16)$. Same training settings are used for DIH without CDA where each corner point of $P_A$ is perturbed at random between $(-16,16)$.

We perform comparison of DSM with a feature-based (FEAT)~\cite{brown2007automatic} and DIH~\cite{detone2016deep} methods. FEAT extract SURF features from a pair of images and performs an exhaustive search for feature matching to estimate the homography. We report the mean residual error (as detailed in~\cite{tella2018probabilistic}) between the GT and estimated relative homographies for SYN (the only sequence with known GT homographies). For quantitative evaluation, we report the average Root Mean Square Error (RMSE) between pair of image patches with known GT homograpies obtained from data augmentation, and average pixel-wise photometric error computed by taking the L1-distance between frame F$_{k+1}$ and reprojected F$_k$ using the estimated homography. We also report qualitative results through visualization.
\section{Results and Discussion}
\begin{figure}[t!]
	\begin{subfigure}[b]{0.2\textwidth}
		\centering
		\includegraphics[width=1.0\textwidth]{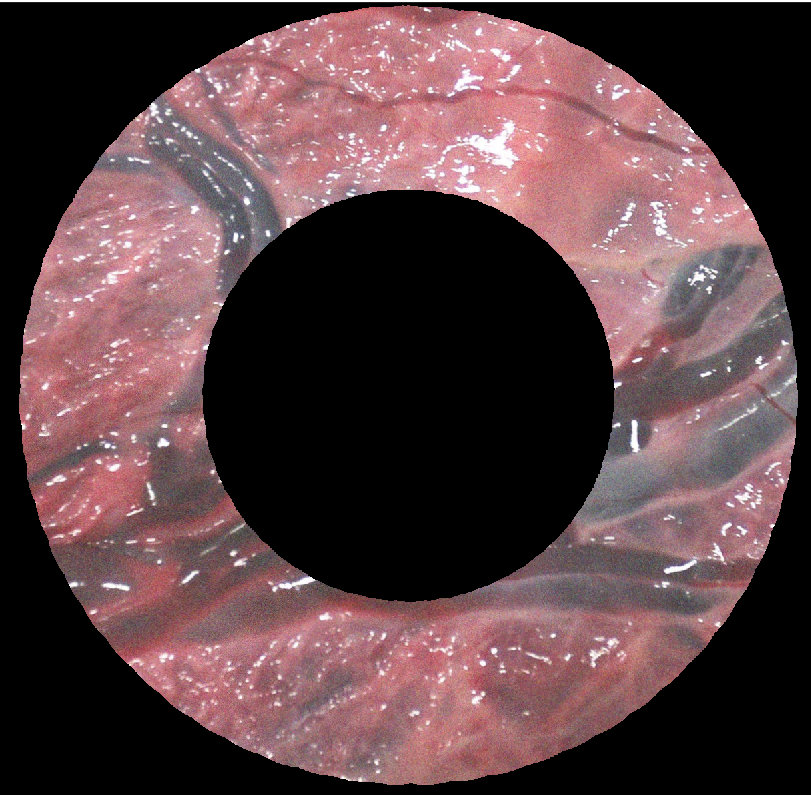}
		\caption{ GT}	
	\end{subfigure}		
	\begin{subfigure}[b]{0.2\textwidth}
		\centering
		\includegraphics[width=0.88\textwidth]{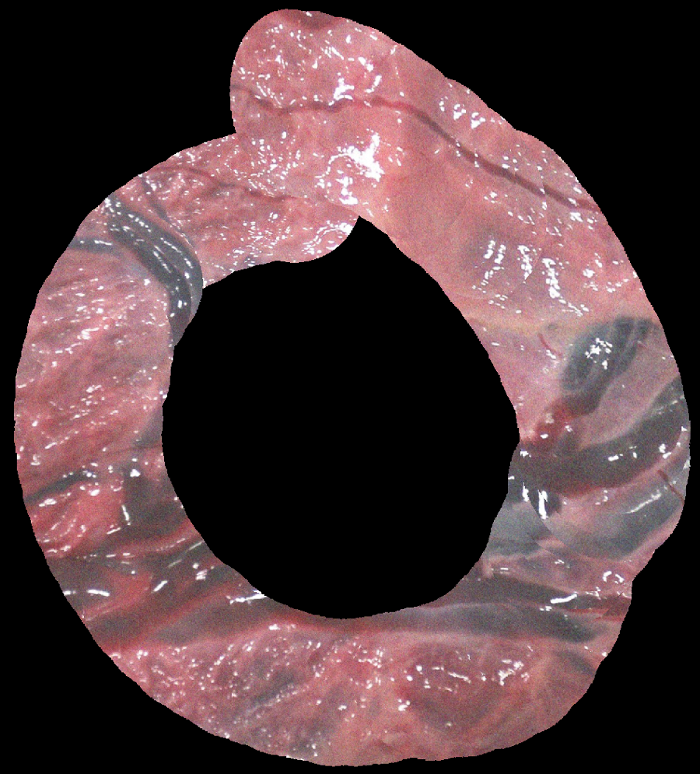}
		\caption{FEAT}					
	\end{subfigure}		
	\begin{subfigure}[b]{0.2\textwidth}
		\includegraphics[width=0.925\textwidth]{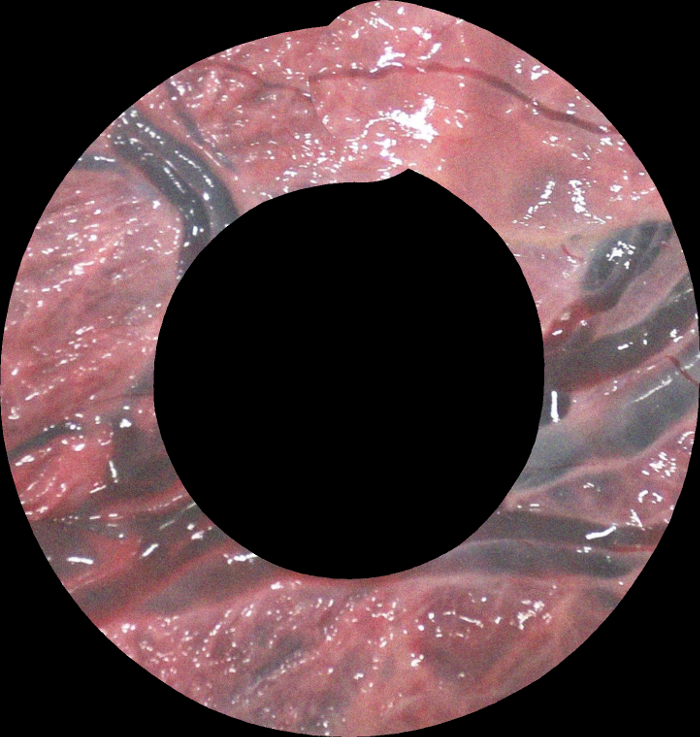}
		\caption{DSM}				
	\end{subfigure}	
	\begin{subfigure}[b]{0.335\textwidth}
		\centering
		\includegraphics[width=0.978\textwidth]{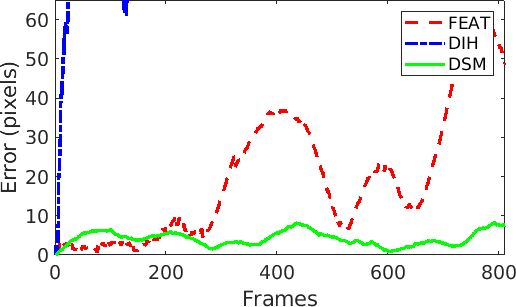}
		\caption{Mean residual error}					
	\end{subfigure}	
	\caption{(a-c) Visualization of mosaics for one circular loop (360 frames) of the SYN sequence. (d) Quantitative comparison of FEAT, DIH and DSM.}
	\label{fig:syn_visualization}     
\end{figure}
\begin{figure}[t!]
\centering
	\begin{subfigure}[b]{0.48\textwidth}
		\centering
		\includegraphics[width=0.9\textwidth]{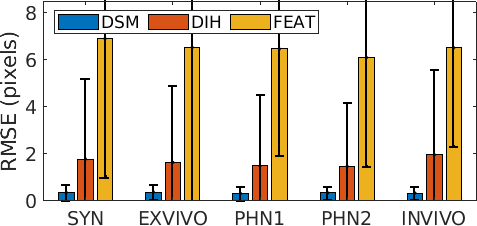}
		\caption{Average RMSE}	
	\end{subfigure}		
	\begin{subfigure}[b]{0.48\textwidth}
		\centering
		\includegraphics[width=0.9\textwidth]{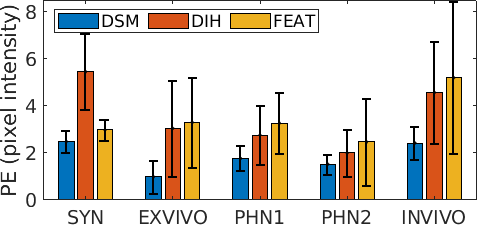}
		\caption{ Average Photometric Error (PE)}			
	\end{subfigure}		
	\caption{Quantitative evaluation and comparison on five diverse fetoscopic videos.}
	\label{fig:rmse_photo}      
\end{figure}

The visualization and comparison results on one circular loop (360 frames) of the SYN sequence are shown in Fig.~\ref{fig:syn_visualization}(a)-(c). Note the small drift in DSM compared to FEAT. Similar behavior is observed from the mean residual error in Fig.~\ref{fig:syn_visualization}(d) where the errors are reported for FEAT, DIH and DSM for the complete length of the sequence (811 frames). It can be seen that the error for FEAT starts increasing after approximately 300 frames and the mosaic starts drifting away. DIH error explodes within a few frames due to the random perturbation during training (Sec.~\ref{sec:deep_homo}). On the other hand, the error for DSM is very small and remains bounded. This is further verified from the low RMSE~(0.36) and photometric~(2.48) errors for DSM (Fig.~\ref{fig:rmse_photo}). Comparison of our proposed DSM with FEAT and DIH is presented in Fig.~\ref{fig:rmse_photo}. Overall the pairwise homography errors are high for FEAT for all five sequences due to poor visual quality and lack of texture in the fetoscopic videos. The RMSE and photometric errors for DIH are low compared to FEAT but are always higher compared to DSM (e.g. RMSE on EX for DIH (1.64) and DSM (0.38)). In DIH, this error accumulated over time during mosaic generation and resulted in a large drift. For EX, PHN1, PHN2 and INVI sequences, the average RMSE errors are 0.38, 0.32, 0.35 and 0.34, and photometric errors are 0.98, 1.76, 1.52, 2.42, respectively. 

\begin{figure}[!t]
	\centering
	\includegraphics[width=0.9\textwidth]{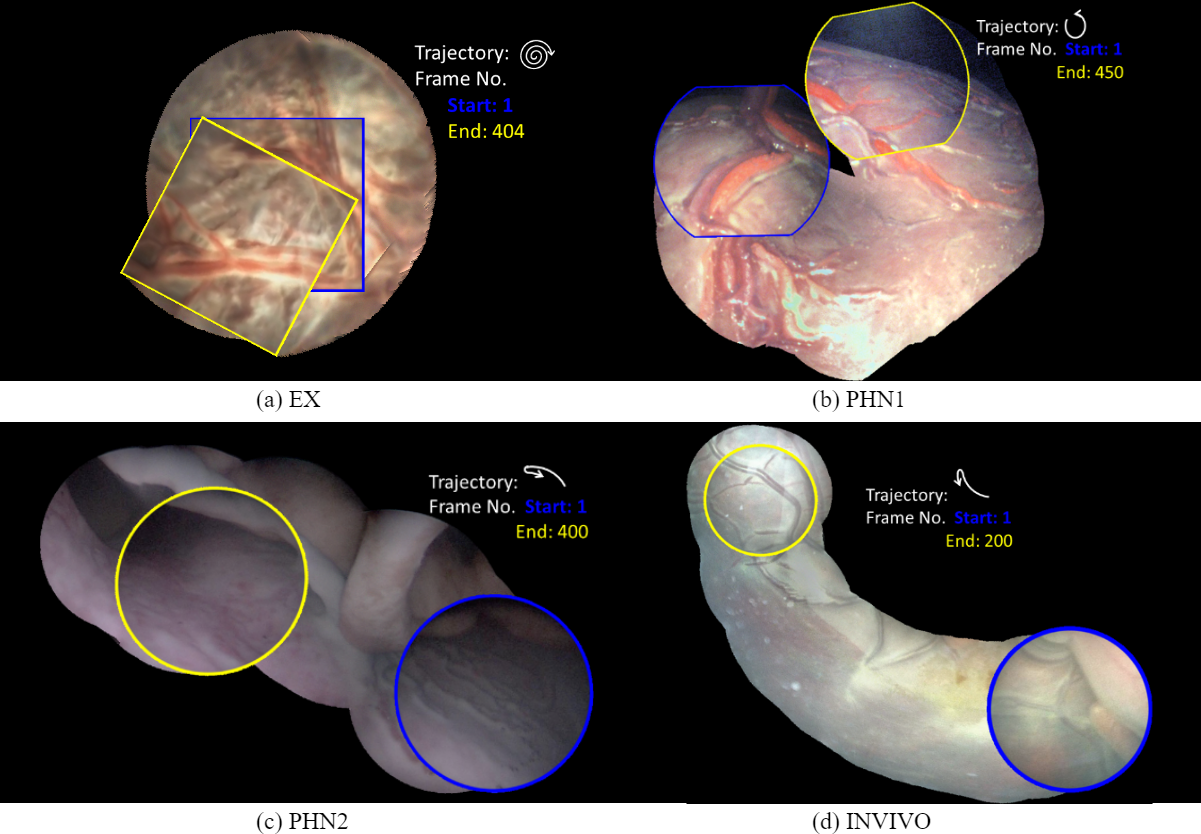}	
	\caption{Qualitative results of the proposed DSM on four diverse fetoscopic videos. The motion trajectories, start and end frames are marked for visualization.}
	\label{fig:qualitative}       
\end{figure}

Mosaics generated using the proposed DSM for the EX, PHN1, PHN2 and INVI sequences are shown in Fig.~\ref{fig:qualitative}. These mosaics are best assessed in the \href{https://youtu.be/wKqXATVl7o8}{supplemental video} that shows the qualitative comparison with respect to FEAT and DIH. DSM created a meaningful mosaic for EX (unseen data) with minimum drift accumulation over time which can be observed from the start and end frames in Fig.~\ref{fig:qualitative}(a). PHN1 contained non-planar views without occlusions with a freehand circular trajectory. DSM generated reliable mosaics with minimum drift (Fig.~\ref{fig:qualitative}(b)), however FEAT drifted away due to non-planar views, insufficient feature matches and long-range videos. PHN2 and INVI represent the most challenging scenarios containing highly non-planar views with heavy occlusions, low resolution and texture paucity. We observe from Fig.~\ref{fig:qualitative}(c)(d) that although the generated mosaics can serve well for increasing the FoV, yet there is a noticeable drift due to highly challenging conditions. Such errors may be corrected by end-to-end training using the photometric loss~\cite{nguyen2018unsupervised}.\looseness=-1

The experimental results show that DSM is capable of handling varying visual quality (varying illumination, specular highlights and low resolution), planar and non-planar views with heavy occlusions. Qualitative evaluation on the unseen EX dataset verified the robustness and generalization capabilities of the proposed DSM. 
Unlike the existing methods that use external sensors for minimizing the drift~\cite{tella2018probabilistic}, DSM relied only on image data and generated meaningful mosaics with minimum drift even for non-planar sequences. 

\noindent{\textbf{Acknowledgments.}} This work was supported through an Innovative Engineering for Health award by Wellcome [WT101957]; Engineering and Physical Sciences Research Council (EPSRC) [NS/A000027/1]. It was additionally supported by the Wellcome/EPSRC Centre for Interventional and Surgical Sciences (WEISS) at UCL [203145Z/16/Z] and EPSRC [EP/N027078/1, EP/P012841/1, EP/P027938/1, EP/R004080/1].

\section{Conclusion}
We proposed a deep sequential mosaicking method for fetoscopic videos acquired through various sources which to our knowledge is a first. Our approach used an existing deep image homography network as a backbone for training but performed controlled data augmentation by assuming that there is only a small change in rotation and translation between two consecutive frames. Due to the lack of texture in fetoscopic sequences, varying specular highlights and turbid amniotic fluid, homography estimation varies between consecutive frames when selecting patch location randomly during testing. To overcome this problem, we proposed an outlier rejection step to obtain a reliable prediction in the least squares sense. Experimental evaluation on five diverse fetoscopic sequences showed that, unlike existing methods that drift rapidly in just a few frames, our method produced mosaics with less drift even for long-range sequences. 

\bibliographystyle{splncs04}
\bibliography{SBbibliography}

\end{document}